\documentclass{article}
\usepackage{latexsym}
\usepackage{amsfonts}
\usepackage{amsmath}
\begin{document}
\pagenumbering{arabic}
\title{Generating Static Black Holes in Higher Dimensional Space-Times}
\author{Emanuel Gallo \thanks
 {egallo@fis.uncor.edu}}
\date{\textit{\small Facultad de Matem\'atica, Astronom\'{\i}a y F\'{\i}sica\\ Universidad Nacional de C\'ordoba, Ciudad Universitaria\\5000 C\'ordoba, Argentina }}
\maketitle

\begin{abstract}
In this article we extend to higher dimensional space-times a
recent theorem proved by Salgado~\cite{salgado} which
characterizes a three-parameter fa\-mily of static and spherically
symmetric solutions to the Einstein Field Equations. As it happens
in four dimensions, it is shown that the Schwarz\-schild,
Reissner-Nordstrom and global monopole solutions in higher
dimensions are particular cases from this family.
\end{abstract}

\section{Introduction}

Recently, Salgado~\cite{salgado} proved a simple theorem
characterizing static sphe\-rically symmetric solutions to the
Einstein's field equations in four dimensions when certain conditions on the energy-momentum tensor are imposed.\\
This theorem allows us to find exact solutions like black-holes
with
diffe\-rent matter fields.\\
The solutions depend on three parameters (one of these being the
cosmolo\-gical constant). It can be easily shown that the
Schwarzschild-de Sitter/An\-ti-de Sitter (SdS/SAdS) or
Reissner-Nordstrom (RN) black-holes are particular
cases from this fa\-mily.\\
These results were independently obtained and used by
Kiselev~\cite{kise1,kise2} in the study of quintessence fields in
black holes and dark matter.\\Similar results were obtained by
{Giamb\'o}~\cite{giambo} in his study of anisotropic
generalization of de Sitter space-time and by
Dymnikova~\cite{dymni} in the study
of a cosmological term as a source of mass.\\
  On the other hand, the study of
solutions of Einstein's equations in higher dimensional
space-times has been very intense for almost
 a decade due to the request of extra dimensions for many
 physics theories (String theory, M theory, Brane-Worlds).\\
 In particular, Myers and Perry~\cite{myers} have found solutions to Einstein's
 equations representing black holes like Schwarzschild,
 Reissner-Nordstrom and Kerr in D-dimensions. Dianyan~\cite{dianyan} has extended
 this work at SdS/SAdS and RN-dS, and  Liu \& Sabra~\cite{sabra} have studied
 general charged configurations in D-dimensional (A)dS spaces with relevant results to this work.\\
 Other examples from solutions to the Einstein's equations in higher dimensions are the analysis of spherically
 symmetric perfect fluids~\cite{krori}, or the collapse of different fluids~\cite{ghosh1,ghosh2,ghosh3} or the study of
 wormholes~\cite{das,cataldo}.\\
 It is therefore interesting to characterize solutions to
 Einstein field equations in higher dimensions than four.\\
 In this paper, we extend the theorem proved by Salgado to
 D-dimensional space-times, where SdS/SAdS and
 RN-dS are particular cases.\\
 In section 2 we briefly review the RN-(dS/AdS) black holes in
 higher dimensions (for a more detailed study see~\cite{dianyan}).\\
 In  section 3 we formulate the extension of the Salgado
 theorem to D-dimensions and finally in  section 4, we will characterize the RN-(dS/AdS) and global monopole solutions
 in this family.

 \section{Reissner-Nordstrom (dS/AdS) Black Holes in Higher Dimensions}

We begin by writing the Einstein-Maxwell equations in
D dimensions with a cosmological constant $\Lambda$.\\
From the Einstein-Maxwell action in D dimensions
\begin{equation}
S=\int d^Dx\sqrt{\mid g \mid}\left\{
R-2\Lambda+\frac{\kappa}{8\pi} F_{ab}F^{ab}\right\},
\end{equation}
where
\begin{eqnarray}
\kappa&=&\frac{8\pi G}{c^4},\\
 F_{ab}&=&A_{a ; b}-A_{b ; a},
\end{eqnarray}
we obtain the following Einstein-Maxwell equations
\begin{eqnarray}
&R_{ab}&-\frac{1}{2}g_{ab}R+\Lambda g_{ab}=\frac{\kappa}{4\pi}\left\{ F_{a}^cF_{bc}-\frac{1}{4}g_{ab}F_{cd}F^{cd}\right\},\\
&F_{a;c}^c&=0,\\
&F_{ab;c}&+F_{bc;a}+F_{ca;b}=0.
\end{eqnarray}
Now, let us consider static and spherically symmetric solutions
from
these equations (We will use units where $G=c=1$).\\
The most general static and spherically symmetric metric  in D
dimensions reads:
\begin{equation}
ds^2=-N^2(r)dt^2+A^2(r)dr^2+r^2d\Omega^2_{D-2},
\end{equation}
where
\begin{equation}
d\Omega^2_{D-2}=d\theta^2_1+\sin^2\theta_1d\theta^2_2+...+\prod_{n=1}^{D-3}\sin^2\theta_nd\theta^2_{D-2}.
\end{equation}
The only non trivial components from $F_{ab}$ are:
\begin{equation}
F_{tr}=-F_{rt}=\frac{Q}{r^{D-2}},
\end{equation}
where $Q$ represents an isolated point charge.\\
Solving the Einstein-Maxwell equations for the metric of eq.(7),
we find ($D>3$)
\begin{equation}
 N^2(r)=\frac{1}{A^2(r)}= 1-\frac{2M}{r^{D-3}}+\frac{2
 Q^2}{(D-3)(D-2)r^{2(D-3)}}-\frac{2\Lambda
 r^2}{(D-2)(D-1)}\\
 \end{equation}
with $M$ a constant of integration.\\
This metric is a generalization to
D dimension of Reissner-Nordstrom-dS/AdS, according to the sign of $\Lambda$.\\
For example, if we make $Q=0$ , $\Lambda > 0$, we obtain the
Schwarschild-dS black hole in D-dimensions. Its horizons were studied by Dianyan~\cite{dianyan}. \\
We will see in the next section (extending the Salgado's theorem)
that these solutions are contained in a more general class of
metrics.

\section{Generating Static Black Holes}
Now, following  Salgado, we state and prove the following theorem:

\newtheorem{theorem}{Theorem}\begin{theorem} Let $(M,g_{ab})$ be a D-dimensional space-time  with
$sign(g_{ab})=D-2$, $D\geq3$, such that:
 (1) it is static and spherically symmetric, (2) it  satisfies the Einstein field equations,
(3) the energy-momentum tensor is given by $T^a_b=T^a_{[f]
b}-\frac{\Lambda}{8\pi}\delta^a_b$, where $T^a_{[f] b}$ is the
energy-momentum tensor of the matter fields, and $\Lambda$ is a
cosmological constant. (4) in the radial gauge coordinate system
adapted to the symmetries of the space-time where
$ds^2=-N^2(r)dt^2+A^2(r)dr^2+r^2d\Omega^2_{D-2}$, the
energy-momentum tensor satisfies the conditions   $T^t_{[f]
t}=T^r_{[f] r}$ and $T^{\theta_1}_{[f] \theta_1}=\lambda T^r_{[f]
r}$ ($\lambda=const \in\mathbb{R})$, (5) it possesses a regular
Killing horizon or a regular origin. Then, the metric of the
space-time is given by
\begin{equation}
ds^2=-\left\{1-\frac{2m(r)}{r^{D-3}}\right\}dt^2+\left\{1-\frac{2m(r)}{r^{D-3}}\right\}^{-1}dr^2+r^2d\Omega^2_{D-2},
\end{equation}
where
\begin{equation}
m(r)=\left\{\begin{array}{rlll} &M+\frac{\Lambda
r^{D-1}}{(D-2)(D-1)} & \mbox{if}\:\:
C=0&\\\\
&M +\frac{\Lambda r^{D-1}}{(D-2)(D-1)}-\frac{8\pi C r^{(D-2)\lambda +1}}{(D-2)[(D-2)\lambda +1]} &{\mbox{if}\:\:\lambda\neq -\frac{1}{D-2}}\:\:;\:\: C\neq 0&\\    \\
 &M +\frac{\Lambda r^{D-1}}{(D-2)(D-1)}-\frac{8\pi C \ln (r)}{D-2} &\mbox{if}\:\:
 \lambda =
 -\frac{1}{D-2}\:\:;\:\: C\neq 0&
\end{array}\\
\right.
\end{equation}
\begin{equation}
T^a_{[f]b}=\frac{C}{r^{(D-2)(1-\lambda)}}diag[1,1,\lambda,...,\lambda],
\end{equation} with $M$ and $C$ integration constants fixed by boundary conditions and fundamental constant of the underlying
matter.
\end{theorem}
\textbf{Proof.} The proof follows exactly the same steps than
Salgado's proof. From the hypothesis 1, we can write the metric in
the following way
\begin{equation}
ds^2=-N^2(r)dt^2+A^2(r)dr^2+r^2d\Omega^2_{D-2}.
\end{equation}
Introducing this metric into Einstein's equations (hypothesis 2)
\begin{equation}
R_{ab}-\frac{1}{2}g_{ab}R=8\pi T_{ab},
\end{equation}
we obtain the following equations for $A(r)$ and $N(r)$
\begin{eqnarray}
\frac{\partial_rA}{Ar}+\frac{(D-3)(A^2-1)}{2r^2}=-\frac{8\pi A^2
}{D-2}(T^t_{[f] t}-\frac{\Lambda}{8\pi}),\\
-\frac{\partial_rN}{Nr}+\frac{(D-3)(A^2-1)}{2r^2}=-\frac{8\pi A^2
}{D-2}(T^r_{[f] r}-\frac{\Lambda}{8\pi}),
\end{eqnarray}
and from eq.(14) and eq.(15) we have that if $a \neq b$ then
\begin{equation}
T^a_b=0,
\end{equation} and  \begin{equation}
T^{\theta_{D-2}}_{\theta_{D-2}}=...=T^{\theta_2}_{\theta_2}=T^{\theta_1}_{\theta_1},
\end{equation}
Rewriting  $A(r)$ as
\begin{equation}
A(r)=\left\{1-\frac{2m(r)}{r^{D-3}}\right\}^{-1/2},
\end{equation}
eq. (16) reads
\begin{equation}
\partial_rm=-\frac{8\pi r^{D-2}}{D-2}(T^{t}_{[f]
t}-\frac{\Lambda}{8\pi}).
\end{equation}
Moreover, subtracting eq. (16) from (17) we have
\begin{equation} \frac{\partial_r(AN)}{AN}=-\frac{8\pi A^2r
}{D-2}(T^t_{[f]t}-T^r_{[f]r}),
\end{equation}
and using the hypothesis 4, ($T^t_{[f]t}-T^r_{[f]r}=0$), we find
that $N=c A^{-1}$, where $c$ is a constant which can be chosen
equal to 1, if
we redefine the time coordinate.\\
 So
\begin{equation}
N(r)=A^{-1}(r)=\left\{1-\frac{2m(r)}{r^{D-3}}\right\}^{1/2},
\end{equation}
with $m(r)$ given by eq. (21).\\
 On the other hand, from the Einstein's equations, it follows that
\begin{equation}
\nabla_a T^a_b=0,
\end{equation}
and this equation in the metric eq.(14) reads
\begin{equation}
\partial_rT^r_{[f]r}=(T^t_{[f]t}-T^r_{[f]r})\frac{\partial_rN}{N}-\frac{D-2}{r}(T^r_{[f]r}-T^{\theta_1}_{[f]\theta_1}).
\end{equation}
Using the hypothesis 4 ($T^t_{[f]t}=T^r_{[f]r}$ ,
$T^{\theta_1}_{[f]\theta_1}=\lambda T^r_{[f]r}$), one obtains
\begin{equation}
\partial_rT^r_{[f]r}=-\frac{D-2}{r}T^r_{[f]r}(1-\lambda),
\end{equation}
and by integration of this equation  we have
\begin{equation}
T^r_{[f]r}=\frac{C}{r^{(D-2)(1-\lambda)}}.
\end{equation}
Finally from eq. (21) we obtain $m(r)$
\begin{equation}
 m(r)=\left\{ \begin{array}{rlll}
&M+\frac{\Lambda r^{D-1}}{(D-2)(D-1)} & \mbox{if}\:\:
C=0&\\\\
&M +\frac{\Lambda r^{D-1}}{(D-2)(D-1)}-\frac{8\pi C r^{(D-2)\lambda +1}}{(D-2)[(D-2)\lambda +1]} &{\mbox{if} \:\:\lambda\neq -\frac{1}{D-2}} \:\:;\:\: C\neq 0&\\    \\
 &M +\frac{\Lambda r^{D-1}}{(D-2)(D-1)}-\frac{8\pi C \ln (r)}{D-2} &\mbox{if}\:\:
 \lambda =
 -\frac{1}{D-2} \:\:;\:\: C\neq 0&
\end{array}\\
\right.
\end{equation}$\diamondsuit$.\\\\
As Salgado remarks, the hypothesis (5) is not used \textit{a
priori}, but it is indeed suggested by the condition
$T^t_{[f]t}=T^r_{[f]r}$
(see ~\cite{salgado}).\\
 Note that if $D=3$, and $\Lambda=C=0$, the solution is a flat
metric, because in three dimensions there is no curvature in
vacuum.\\
We could do a study of the energy condition on the matter fields
as in~\cite{salgado}, but it gives the same results as in four
dimensions, and it not will be repeated here.\\
The only change is on the \textit{effective equation of state} as
measured by an observer staying in rest in the coordinate system
given in the theorem:
\begin{equation}
p_f=\frac{1}{D-1}(T^r_r+\sum^{D-2}_{i=1}T^{\theta_i}_{\theta_i})=-\frac{[1+(D-2)\lambda]}{D-1}\rho,
\end{equation}
where $p_f$ is the  effective pressure and $\rho$ is the
energy-density of the matter defined by
\begin{equation}
\rho=-T^t_t=-T^r_r=-\frac{C}{r^{(D-2)(1-\lambda)}}.
\end{equation}
For example, if we have a electromagnetic field, ($\lambda=-1$,
see the next section) we get
\begin{equation}
p_f=\frac{D-3}{D-1}\rho,
\end{equation}
and if $D=4$ we have the well known state equation $p_f=\rho/3$.

\section{Characterizing RN-(dS/AdS) and global \\monopoles solutions in higher dimensions}
Now, we characterize some known solutions in this three-parameter
family.\\Let us begin with the Reissner-Nordstrom (dS/AdS) black
holes.\\
 For a static spherically symmetric electrical field we have
 \begin{equation}
 F_{tr}=-F_{rt}=\frac{Q}{r^{(D-2)}},
 \end{equation}
 and from the electromagnetic energy-momentum tensor
 \begin{equation}
 T_{ab}=\frac{1}{4\pi}(F_{a}^cF_{bc}-\frac{1}{4}g_{ab}F_{cd}F^{cd}),
 \end{equation}
we can see that
\begin{equation}
T^a_b=-\frac{Q^2}{8\pi r^{2(D-2)}}diag[1,1,-1,...,-1].
\end{equation}
Then, we have
 \begin{equation} \lambda=-1,
\end{equation} and
\begin{equation}
C=-\frac{Q^2}{8\pi}.
\end{equation}
 Then, if $D>3$ putting  these values for the parameters into the metric from the theorem,
 we have all RN-(dS/AdS) metrics in D dimensions (eq.10).\\
 For example, if D=4, we can see that $A(r)$ and $N(r)$  reads
\begin{equation}
N(r)=A^{-1}(r)=\left\{ 1-\frac{2M}{r}+\frac{Q^2}{r^2}-
\frac{\Lambda r^2}{3}\right\}^{1/2},
\end{equation}
which is the well-known Reissner-Nordstrom (dS/AdS) metric in four
dimensions.\\
In the case D=3, (using the theorem) we see that the metric for a
charged (dS/AdS) black hole in three dimensions reads
\begin{equation}
ds^2=-\left\{\widetilde{M}-\Lambda r^2-2Q^2\ln
r\right\}dt^2+\left\{\widetilde{M}-\Lambda r^2-2Q^2\ln
r\right\}^{-1}dr^2+r^2d\theta_1^2,
\end{equation}
where $\widetilde{M}=1-2M$.\\ Properties of these solutions and
2+1 black holes in general have
been stu\-died in~\cite{park,kogan,banados1}.\\
Finally, we study another solution  which represents a black hole
with a trivial global
monopole inside.\\
The energy-momentum tensor for a global monopole in D dimensions
is given by
\begin{equation}
T_{ab}=\nabla_a \phi^i \nabla_b \phi_i-\frac{g_{ab}}{2}(\nabla
\phi^i)^2-g_{ab}\frac{\nu}{4}(\phi^i \phi_i -\eta^2)^2.
\end{equation}
Then, if we have a trivial monopole~\cite{vilenkin}
\begin{equation}
\phi^i=\frac{\eta x^i}{r} \qquad (i=1,2,...D-1)
\end{equation}
which remains in the vacuum state
$V(\phi_i\phi^i)=\frac{\nu}{4}(\phi_i\phi^i-\eta^2)^2=0$ it can be
shown that
\begin{equation}
T^a_b=-\frac{(D-2)\eta^2}{2r^2}diag[1,1,\frac{D-4}{D-2},...,\frac{D-4}{D-2}],
\end{equation}
and then we note that this monopole satisfies the
 theorem's conditions on its energy-momentum tensor, with $\lambda=\frac{D-4}{D-2}$ and
$C=-\frac{(D-2)\eta^2}{2}$.\\
If $D>3$ we obtain a D-dimensional solution representing a global
monopole inside a static and spherically symmetric black-hole:
\begin{equation}
N(r)=A^{-1}(r)=\left\{1-\frac{2M}{r^{D-3}}-\frac{2\Lambda r^2}{
(D-2)(D-1)}-\frac{8\pi\eta^2}{(D-3)}\right\}^{1/2}
\end{equation}
For example, if $D=4$, then  $\lambda=0$ and we have (with
$\Lambda=0$) the well-known solution
\begin{equation}
N(r)=A^{-1}(r)=\left\{1-\frac{2M}{r}-8\pi\eta^2\right\}^{1/2}.
\end{equation}
On the other hand, if $D=3$ (i.e $\lambda=-1$) and $\Lambda=0$, we
can see (using the theorem) that the monopole's 3-metric is
\begin{equation}
N(r)=A^{-1}(r)=\left\{\widetilde{M}-8\pi\eta^2\ln(r)\right\}^{1/2},
\end{equation}
with $\widetilde{M}=1-2M$.
\section{Conclusions}
In this article, we have extended to D dimensions a simple theorem
which cha\-racterizes spherically symmetric solutions to the
Einstein field equations under certain conditions of the
energy-momentum
tensor.\\
It can be seen that there exist some matter fields with a
parameter $\lambda$ so that $T^\theta_{[f]\theta}=\lambda
T^r_{[f]r}$, and this parameter characterizes (partially) the
radial dependence of the metric generated by the
matter fields.\\
For a static spherically symmetric electromagnetic field, it was
shown
that $\lambda=-1$ independently of the dimension.\\
On the other hand, for the case of global monopoles solutions,
$\lambda$ depends explicitly on the dimensionality, i.e.,
$\lambda=\frac{D-4}{D-2}$.\\
Finally, it is very interesting to ask if there exist another
matter fields which sa\-tisfies the theorem's conditions in D
dimensions.

\section{Acknowledgements}

The author thanks Carlos Kozameh for the reading and revision of
this paper. The author is supported by CONICET.


\begin{thebibliography}{9}
\bibitem{salgado} { \textrm M. Salgado.}  {\it A simple theorem to generate
exact black-hole solutions.} {\textrm Class.Quantum Grav.
\textbf{20}, 4451 (2003).}
\bibitem{kise1} { \textrm V.V. Kiselev.}  {\it Quintessence and Black Holes.}\\ {\textrm Class.Quantum.Grav. \textbf{20} 1187, (2003)}
\bibitem{kise2} { \textrm V.V. Kiselev.}  {\it Quintessential solution of dark matter rotation curves and its simulation by extra dimensions.
} {\textrm gr-qc/0303031 (2003).}
\bibitem{giambo} { \textrm R. Giamb\'o}  {\it Anisotopic generalization of de Sitter spacetime.} {\textrm Class.Quant.Grav. \textbf{19} 4399, (2002)}
\bibitem{dymni} { \textrm I. Dymnikova}  {\it Cosmological term as a source of mass.}\\ {\textrm Class.Quantum.Grav. \textbf{19} 725,  (2003).}
\bibitem{myers} {  R.C. Myers and M.J. Perry.}  {\it Black Holes in Higher Dimensional Space-times.} {\textrm Ann. Phys. (N.Y.) \textbf{172}, 304(1986)}.
\bibitem{dianyan} {\textrm X. Dianyan.}  {\it Exact solutions of Einstein and Einstein-Maxwell equations in higher-dimensional
spacetime.} {\textrm Class.Quantum Grav. \textbf{5}, 871 (1988).}
\bibitem{sabra} {\textrm J. Liu and W. Sabra.} {\it Charged configurations in (A)dS spaces.}\\ {\textrm hep-th/0307300 (2003).}
\bibitem{krori} {\textrm K. D. Krori, P. Borgohain, and Kanika
Das.
}  {\it Spherically symmetric solutions in higher dimensions.}
{\textrm J. Math. Phys. \textbf{30}, 2315 (1989).}
\bibitem{ghosh1} {\textrm S.G. Ghosh and D.W. Deshkar.
}  {\it Gravitational Collapse of Perfect Fluid in Self-Similar
Higher Dimensional Space-Times.} {\textrm Int.J.Mod.Phys. D
\textbf{12} 913 (2003).}
\bibitem{ghosh2} {\textrm S.G. Ghosh and A. Banerjee.
}  {\it Non-marginally bound inhomogeneous dust collapse in higher
dimensional space-time.} {\textrm Int.J.Mod.Phys. D \textbf{12}
639 (2003).}
\bibitem{ghosh3}
{\textrm K. S.G. Ghosh and N. Dadhich.}  {\it Gravitational
collapse of Type II fluid in higher dimensional space-times.}
{\textrm Phys.Rev. D \textbf{65} 127502 (2002).}
\bibitem{das} {\textrm K. A. Debenedictis and A. Das.}  {\it Higher Dimensional Wormhole Geometries with
Compact Dimensions.} {\textrm Nucl.Phys. B \textbf{653} 279
(2003).}
\bibitem{cataldo}
{\textrm K. M. Cataldo, P. Salgado, and P. Minning. }  {\it
Self--dual Lorentzian wormholes in n--dimensional Einstein
gravity.} {\textrm Phys.Rev. D \textbf{66} 124008 (2002).}
\bibitem{park}
{\textrm D. Park and S. Yang. }  {\it Geodesics motions in 2+1
dimensional charged black holes.} {\textrm Gen.Rel.Grav.
\textbf{31} 1343 (1999).}
\bibitem{kogan}
{\textrm I. Kogan. }  {\it About some exact solutions for 2+1
gravity coupled to gauge fields.} {\textrm Mod.Phys.Lett. A
\textbf{7} 2341 (1992).}
\bibitem{banados1} {\textrm M. Ba\~{n}ados, C.
Teitelboim and J. Zanelli. }  {\it Geometry of the 2+1 black
hole.} {\textrm Phys.Rev. D \textbf{48} 1506 (1993).}
\bibitem{vilenkin} {\textrm A. Vilenkin and E.
Shellard. }  {\it Cosmic Strings and Other Topological Defects.}
{\textrm Canbrige University Press (1994).}
\end{thebibliography}
\end{document}